\documentclass[12pt]{article}
\usepackage{amssymb}
\usepackage{graphicx}
\usepackage{amsmath}
\usepackage{xcolor}

\input{tcilatex}

\begin{document}

\begin{center}
{\Large Finding network effect of randomized treatment under weak
assumptions for any outcome and any effect heterogeneity\medskip }

\begin{tabular}{c}
Myoung-jae Lee \\ 
Department of Economics, Korea University \\ 
Seoul 02841, South Korea,\ myoungjae@korea.ac.kr; \\ 
Department of Finance, Accounting \& Economics \\ 
University of Nottingham Ningbo China \\ 
Ningbo 315100, China%
\end{tabular}%
\bigskip \bigskip \bigskip \bigskip \bigskip
\end{center}

In estimating the effects of a treatment/policy with a network, an unit is
subject to two types of treatment: one is the direct treatment on the unit
itself, and the other is the indirect treatment (i.e., network/spillover
influence) through the treated units among the friends/neighbors of the
unit. In the literature, linear models are widely used where either the
number of the treated neighbors or the proportion of them among the
neighbors represents the intensity of the indirect treatment. In this paper,
we use a nonparametric network-based \textquotedblleft causal reduced form
(CRF)\textquotedblright\ that allows any outcome variable (binary, count,
continuous, ...) and any effect heterogeneity. Then we assess those popular
linear models through the lens of the CRF. This reveals what kind of
restrictive assumptions are embedded in those models, and how the
restrictions can result in biases. With the CRF, we then conduct almost
model-free estimation and inference for network effects.\bigskip \bigskip
\bigskip \bigskip \bigskip 

\textbf{Running Head}: Network effect for any outcome and heterogeneity.


\textbf{Compliance with ethical standard, no conflict of interest, and no AI
usage:}\ no human/animal subject is involved in this research, and there is
no conflict of interest to disclose. Also, no generative-AI-related
technology has been used for this paper.\textbf{\pagebreak }

\section{Introduction}

\qquad Consider $N$ units connected through a network, and a binary
randomized treatment/policy $D$. Let $Y_{i}$ be an outcome of unit $i=1$%
\thinspace $...N$, $F_{i}$\textit{\ }be the number of direct
(\textquotedblleft first-order\textquotedblright ) friends/neighbors of unit 
$i$ (\textquotedblleft $F$\textquotedblright\ for friends), and $T_{i}$\ be
the total number of the treated direct friends/neighbors subject to $%
T_{i}\leq F_{i}$. When treatment spillover is present, $Y_{i}$ is influenced
not only by its own treatment $D_{i}$, but also by the treatments received
by friends: two treatments (direct, and indirect through the network) appear.

\qquad Let $U_{0i}$ and $U_{1i}$ be error terms. Define the treated
proportion among friends $R_{i}\equiv T_{i}/F_{i}$. Two popular
network-effect linear models: with $\beta $ parameters, under $F_{i}>0$,%
\begin{eqnarray}
\ Y_{i} &=&\beta _{0}+\beta _{d}D_{i}+\beta _{\tau }T_{i}+\beta _{f}F_{i}\
+U_{0i};  \TCItag{T-model} \\
Y_{i} &=&\beta _{0}+\beta _{d}D_{i}+\beta _{r}R_{i}+U_{1i}=\beta _{0}+\beta
_{d}D_{i}+\frac{\beta _{r}}{F_{i}}T_{i}+U_{1i}.  \TCItag{R-model}
\end{eqnarray}%
Kim (2025) considered these models, using the terminology \textquotedblleft $%
T$-regression and $\bar{D}$ regression\textquotedblright\ instead of T- and
R-models. See, e.g., Miguel and Kremer (2004), Oster and Thornton (2012),
Bryan et al. (2014), and Cai et al. (2015) for empirical examples using the
above models (or their variants), whose theoretical justifications were
provided by Manski (2013), Aronow and Samii (2017), and Leung (2020).
Henceforth, we often omit the subscript $i$ unless doing so causes confusion.

\qquad In the T- and R-models, $\beta _{d}$ is the direct effect common
across the two models, but the effect (i.e., slope)\ of the network
treatment $T$ (the number of treated friends) differs, due to $T$ being
taken as the network treatment in the second expression of the R-model,
instead of the network treatment $R$ in the first expression. This way of
rewriting the R-model helps comparing the two models. Whereas $D$ is binary,
the network treatment $T$ is non-binary with an integer-valued treatment
intensity. Network effects can be defined and accounted for in various ways
(see, e.g., Hu et al. (2022) and references therein), but the T-, and
R-models postulate that it operates only through the single route $T$ or $R$.

\qquad A problem with the above models is that the interaction $DT$ is ruled
out. A general model encompassing the two models while allowing for
interaction effects is: for an error term $U_{2}$, still under $F>0$,%
\begin{eqnarray}
&&Y_{i}=\beta _{0}+\beta _{f}F_{i}+\beta _{d}D_{i}+\beta _{\tau }T_{i}+\beta
_{r}R_{i}+\beta _{d\tau }D_{i}T_{i}+\beta _{dr}D_{i}R_{i}+U_{2i}  \notag \\
&&\ =\beta _{0}+\beta _{f}F_{i}+\beta _{d}D_{i}+(\beta _{\tau }+\frac{\beta
_{r}}{F_{i}})T_{i}+(\beta _{d\tau }+\frac{\beta _{dr}}{F_{i}}%
)D_{i}T_{i}+U_{2i}.  \TCItag{TR-model}
\end{eqnarray}%
We can see which (or neither) is correct between the T- or R-models by
estimating the TR-model. The total effect of a vaccine consists of the
direct effect of the unit getting vaccinated and the network effect of the
unit's friends getting vaccinated. The indirect effect would be greater if
the unit is also vaccinated, which is the interaction effect of $DT$.

\qquad Although the TR-model is more general, there are questions for all
three models. First, in general, the linear models are invalid when $Y$ is
noncontinuous, and thus for noncontinuous $Y$, is it all right to still
employ the linear models or should a nonlinear outcome model be considered
as in Oster and Thornton (2012)? \ Second, in reality, the network effect of 
$T$ is likely heterogeneous in $F$ (and other covariates), neither constant
nor proportional to $F^{-1}$. Third, in view of this issue, we may want to
control $F$ to find the heterogeneous effect: how exactly then should $F$ be
controlled in this case?

\qquad The issue of controlling $F$ may be viewed more generally as the
issue of specifying the network-effect part as a function of $(T,F)$. The
T-model adopts the additive form $\beta _{\tau }T+\beta _{f}F$, and the
R-model adopts the ratio form $\beta _{r}R=\beta _{r}T/F$. Given that the
discrepancy between two groups $D=0,1$ can be assessed with the linear form $%
E(Y|D=1)-E(Y|D=0)$ or the ratio form $E(Y|D=1)/E(Y|D=0)$, both T-model and
R-models may be plausible. What is right between the additive and ratio
forms might be an empirical matter. Surprisingly, however, we know the true
model.\ This is intuitively explained in the following, using the basic
framework of the effect of only $D$ on $Y$.

\qquad As is usual in treatment effect analysis, let $X$ denote covariates,
and $(Y^{0},Y^{1})$ be the potential outcomes\ for $D=0,1$, so that the
observed outcome is $Y=Y^{0}+(Y^{1}-Y^{0})D$. Since we deal with a
randomized $D$, `$(Y^{0},Y^{1})\amalg D|X$' trivially holds, where $\amalg $
stands for independence. Then, take $E(\cdot |D,X)$ on $%
Y=Y^{0}+(Y^{1}-Y^{0})D$ to obtain:%
\begin{equation*}
E(Y|D,X)=E(Y^{0}|X)+E(Y^{1}-Y^{0}|X)D.
\end{equation*}%
Now, defining $U\equiv Y-E(Y|D,X)$ renders a nonparametric \textquotedblleft
causal reduced form (CRF)\textquotedblright\ that holds for any $Y$ (binary,
count, continuous, ...):%
\begin{equation}
Y=E(Y^{0}|X)+E(Y^{1}-Y^{0}|X)D+U,\ \ \ \ \ E(U|D,X)=0.  \tag{1.1}
\end{equation}

\qquad This CRF is the true model that holds always, where $E(Y^{0}|X)$ is
the $X$-conditional intercept and $E(Y^{1}-Y^{0}|X)$ is the $X$-conditional
slope of $D$. The name \textquotedblleft CRF\textquotedblright\ is
appropriate because it is a derived/reduced form (RF), not a structural form
(SF), and yet it contains the causal parameter of interest $E(Y^{1}-Y^{0}|X)$%
. Of course, we do not know the functional forms of $E(Y^{0}|X)$ and $%
E(Y^{1}-Y^{0}|X)$, but we can see the assumptions to justify the usual
linear model, say $Y=\beta _{x}^{\prime }X+\beta _{d}D+error$: $%
E(Y^{0}|X)=\beta _{x}^{\prime }X$ and $E(Y^{1}-Y^{0}|X)=\beta _{d}$. If $X$
is \textquotedblleft rich/detailed\textquotedblright , then $%
E(Y^{0}|X)=\beta _{x}^{\prime }X$ might be plausible, but it would be
difficult to justify $E(Y^{1}-Y^{0}|X)=\beta _{d}$ because treatment effects
are almost always heterogeneous in reality---think of the Covid vaccine
effects. In view of this, the CRF may be estimated by ordinary least squares
(OLS) estimator under $E(Y^{0}|X)=\beta _{x}^{\prime }X$ and $%
E(Y^{1}-Y^{0}|X)=\beta _{dx}^{\prime }X$, which yields $Y=\beta _{x}^{\prime
}X+\beta _{dx}^{\prime }XD+error$.

\qquad Turning back to the network effect framework, the situation is more
complex due to the two types of treatment: the binary (direct) treatment $%
D=0,1$ on each unit, and an integer (indirect) treatment $T=0,1...F$. The
goals of this paper are establishing a nonparametric CRF appropriate for
this situation to show what kind of restrictions are (unknowingly) embedded
in the above linear models, and explaining how to estimate all effects of
interest using the CRF derived by doing analogously to deriving (1.1).

\qquad In the remainder of this paper, Section 2 derives two versions of
CRF, and makes various remarks on the CRF's. Section 3 provides a simulation
study. Finally, Section 4 concludes this paper.

\section{Causal Reduced Form (CRF) for Network Effect}

\qquad Let $Y_{i}(d_{1},d_{2}...d_{N})$ be the potential outcome of unit $i$
for a treatment vector $(d_{1},d_{2}...d_{N})$ for all $N$ units in the
network. Under some assumptions, $Y_{i}(d_{1},d_{2}...d_{N})$\textit{\
depends only on }$(d_{i},t_{i})$\textit{\ where }$t_{i}$\textit{\ is the
realized value of }$T_{i}$\textit{. We maintain such assumptions throughout
this paper}; e.g., Assumption 2 (with $g_{i}=t_{i}$) in Forastiere et al.
(2021), or more specifically, the assumption that network effect occurs only
among the direct friends who are `exchangeable' in Leung (2020).

\subsection{Unrestricted CRF}

\qquad Our causal effects of interest conditional on $F$ are:%
\begin{equation}
\delta _{t}(F)\equiv E(Y^{1t}-Y^{0t}|F),\ \tau _{1t}(F)\equiv
E(Y^{1t}-Y^{10}|F),\ \tau _{0t}(F)\equiv E(Y^{0t}-Y^{00}|F).  \tag{2.1}
\end{equation}%
Here, $\delta _{t}(F)$ is the direct (thus \textquotedblleft
delta\textquotedblright ) effect of the own treatment, when the unit has $t$
treated among a total of $F$ friends; $\tau _{1t}(F)$ is the network effect
on a treated unit of having $t$ treated out of $F$; and $\tau _{0t}(F)$ is
the network effect on a control unit of having $t$ treated out of $F$.

\qquad Define the (\textquotedblleft net\textquotedblright ) interaction
effect of $(D=1,T=t)$ as (Choi and Lee, 2018):%
\begin{equation*}
\Delta Y_{t}^{\pm }\equiv
Y^{1t}-Y^{10}-Y^{0t}+Y^{00}=Y^{1t}-Y^{00}-(Y^{10}-Y^{00})-(Y^{0t}-Y^{00})
\end{equation*}%
which is the \textquotedblleft gross\textquotedblright\ interaction effect\ $%
Y^{1t}-Y^{00}$ of $(D=1,T=t)$ minus the \textquotedblleft
partial\textquotedblright\ effects\ $Y^{10}-Y^{00}$ of $D=1$ and$\
Y^{0t}-Y^{00}$ of $T=t$. Let $1[A]\equiv 1$ if $A$ holds and $0$ otherwise.
Theorem 1 below is proved in the appendix.\bigskip

\textbf{THEOREM 1.}\textit{\ The following CRF holds for any }$Y$\textit{\
(binary, count, continuous,...): with }$E(U|D,T,F)=0$\textit{\ by
construction,}%
\begin{eqnarray}
Y &=&E(Y^{00}|F)+\delta _{0}(F)D+\sum_{t=1}^{F}\tau
_{0t}(F)1[T=t]+\sum_{t=1}^{F}\tau _{\pm t}(F)\cdot D1[T=t]+U,  \notag \\
\tau _{\pm t}(F) &\equiv &E(Y^{1t}-Y^{10}-Y^{0t}+Y^{00}|F)=\tau
_{1t}(F)-\tau _{0t}(F)=\delta _{t}(F)-\delta _{0}(F);  \TCItag{CRF1}
\end{eqnarray}%
\textit{non-zero interaction effect is }$\tau _{\pm t}(F)\neq
0\Longleftrightarrow \tau _{1t}(F)\neq \tau _{0t}(F)\Longleftrightarrow
\delta _{t}(F)\neq \delta _{0}(F)$\textit{.\bigskip }

\textbf{Remark 1.} As $T=\sum_{t=1}^{F}t\cdot 1[T=t]$, in view of CRF1, the
restrictions in the linear models are:%
\begin{equation}
\begin{tabular}{lcccc}
\hline\hline
& $E(Y^{00}|F)$ & $\delta _{0}(F)$ & $\tau _{0t}(F)$ & $\tau _{\pm t}(F)$ \\ 
\hline
T-model: & $\beta _{0}+\beta _{f}F$ & $\beta _{d}$ & $\beta _{\tau }t$ & $0$
\\ 
R-model: & $\beta _{0}$ & $\beta _{d}$ & $(\beta _{r}/F)t$ & $0$ \\ 
TR-model: & $\beta _{0}+\beta _{f}F$ & $\beta _{d}$ & $(\beta _{\tau }+\beta
_{r}/F)t$ & $(\beta _{d\tau }+\beta _{dr}/F)t$ \\ \hline\hline
\end{tabular}
\tag{2.2}
\end{equation}%
Some restrictions here are strong; e.g., $E(Y^{00}|F)=\beta _{0}$ in the
R-model, and $\delta _{0}(F)=\beta _{d}$ in all three models. Also, $\tau
_{0t}(F)$\textit{\ and }$\tau _{\pm t}(F)$\textit{\ are unknown functions of 
}$(t,F)$\textit{, and yet the specifications of these functions are not even
linear in }$(t,F)$\textit{\ in all three-models.}\bigskip 

\textbf{Remark 2.} All effects in CRF1 are allowed to be heterogeneous in $F$%
, but since $F$ is integer-valued, those effects can be specified as linear
in the dummy variables for $F$ without loss of generality: e.g., with $f=0,1$%
\thinspace $...\bar{f}$ for an upper bound $\bar{f}$ on $F$ with $P(F=\bar{f}%
)>0$,%
\begin{equation*}
\delta _{0}(F)=\sum_{f=0}^{\bar{f}}\delta _{0}(f)1[F=f]=\delta
_{0}(0)+\sum_{f=1}^{\bar{f}}\{\delta _{0}(f)-\delta _{0}(0)\}1[F=f].
\end{equation*}%
Hence, OLS\ of $Y$ on $(1,D,1[T=t],D1[T=t])$ interacted with $(1[F=1]...1[F=%
\bar{f}])$ provides nonparametric estimators for $\{E(Y^{00}|F),\delta
_{0}(F),\tau _{0t}(F),\tau _{\pm t}(F)\}$, from which we can then identify $%
\tau _{1t}(F)$ using $\tau _{1t}(F)=\tau _{0t}(F)+\tau _{\pm t}(F)$ in
Theorem 1, and $\delta _{t}(F)$ using $\delta _{t}(F)=\delta _{0}(F)+\tau
_{\pm t}(F)$. Instead of this \textquotedblleft long OLS\textquotedblright ,
we may as well do the \textquotedblleft short OLS\textquotedblright\ of $Y$
on $(1,D,1[T=t],D1[T=t])$ separately for each subsample $F=f$.\bigskip 

\textbf{Remark 3. }\textquotedblleft CRF\textquotedblright\ may sound
strange, but CRF has been fruitfully used in various contexts recently:\ Lee
(2018, 2021, 2024), Mao and Li (2020), Choi et al. (2023), Lee et al.
(2023), Lee and Han (2024), Kim and Lee (2024), Lee et al. (2025), and Kim
(2025). Traditionally, often a tightly specified SF is used that contains a
parameter of interest. Instead of this, \textquotedblleft targeted
learning\textquotedblright\ (Van der Laan and Rose, 2011) declares a causal
parameter of interest first, and then considers how to identify and estimate
it. Our CRF-based approach stands on a middle ground, in the sense that it
uses a nonparametric RF with causal parameters.\bigskip 

\textbf{Remark 4.} Continuing on the CRF literature, CRF with network effect
has been derived already by Kim (2025). Nevertheless, CRF1 and the CRF in
Kim (2025, Appendix, equation (8)) differ in that we focus on the
\textquotedblleft network level effect\textquotedblright\ of $T=t$ relative
to the baseline $T=0$, whereas Kim (2025) focuses on the \textquotedblleft
network change effect\textquotedblright\ of $T=t$ relative to $T=t-1$. The
difference further leads to a difference in the interaction effect:%
\begin{eqnarray}
&&\ E(Y^{1t}-Y^{10}-Y^{0t}+Y^{00}|F)\text{ \ \ for \ \ }t\leq T\leq F\text{\
\ \ in CRF1;}  \TCItag{2.3} \\
&&\sum_{\tau =1}^{F}1[t\leq T]\cdot E(Y^{1t}-Y^{1,t-1}-Y^{0t}+Y^{0,t-1}|F)%
\text{ \ \ in \ \ Kim (2025).}  \TCItag{2.4}
\end{eqnarray}%
Also, the CRF proofs differ, as they were developed independently of each
other. The interaction effect (2.3) that appeared first in Choi and Lee
(2018) for binary $T$ is simpler than (2.4), but \textquotedblleft
telescoping\textquotedblright\ reveals that the two effect definitions are
the same:%
\begin{eqnarray*}
&&(2.4)=\sum_{t=1}^{F}1[t\leq T]\cdot E(Y^{1t}-Y^{1,t-1}-Y^{0t}+Y^{0,t-1}|F)
\\
&&=E(Y^{1t}-Y^{1,t-1}-Y^{0t}+Y^{0,t-1}|F)+E(Y^{1,t-1}-Y^{1,t-2}-Y^{0,t-1}+Y^{0,t-2}|F)+\ ...
\\
&&\ \ \ \ +E(Y^{11}-Y^{10}-Y^{01}+Y^{00}|F)\ =\
E(Y^{1t}-Y^{0t}-Y^{10}+Y^{00}|F)=(2.3).
\end{eqnarray*}

\textbf{Remark 5.} We prefer (2.3) to (2.4) for its simplicity, and there is
yet another reason:\ $E(Y^{1t}-Y^{10}-Y^{0t}+Y^{00}|F)$ in (2.3) is more
readily applicable to non-ordered treatments. E.g., if $T$ represents
non-ordered categories, then (2.3) is the interaction effect of category $T=t
$ relative to the fixed base category $T=0$, whereas (2.4) adds up\ the
different category interaction effects of $T=t$ relative to the varying base
category $T=t-1$, which is an unnecessary complication for non-ordered
treatments. Non-ordered network treatments can occur, if there are different
types of friends; e.g., having 2 close friends may not be ordered relative
to having 10 not-so-close friends.

\subsection{Restricted CRF with Power-Function Specifications}

\qquad The aforementioned long OLS is cumbersome if the support of $F$ is
large, and is numerically unstable if $P(F=f)$ is too small for some $f$.
E.g., suppose $\bar{f}=20$ and the maximum value of $T$ is $10$. Then the
long OLS\ would have about $20+20+20\times 10+20\times 10=440$ parameters,
where the four parts correspond to those in CRF1. Here, we reduce the long
OLS\ dimension with power-function approximations in $t$ and $F$.\bigskip

\textit{ASSUMPTION 1. For a known }$J$, \textit{and some }$\beta $, $\delta $
and $\tau $ \textit{parameters,}%
\begin{equation*}
E(Y^{00}|F)=\sum_{j=0}^{J}\beta _{0j}F^{j},\ \ \delta
_{0}(F)=\sum_{j=0}^{J}\delta _{0j}F^{j},\ \ \tau _{0t}(F)=\sum_{j=0}^{J}\tau
_{01j}F^{j}t,\ \ \tau _{\pm t}(F)=\sum_{j=0}^{J}\tau _{\pm 1j}F^{j}t.
\end{equation*}

\qquad Assumption 1 specifies order-$J$ power-functions for $E(Y^{00}|F)$
and $\delta _{0}(F)$, and order-$J$ (in $F$) and order-$1$ (in $t$) power
functions for $\tau _{0t}(F)$ and $\tau _{\pm t}(F)$. Of course, a
higher-order can be used for $t$, if desired. Using $T=\sum_{t=1}^{F}t1[T=t]$%
, Assumption 1 simplifies CRF1 much by turning the $1[T=t]$ and $D1[T=t]$
parts in CRF1 into $(\sum_{j=0}^{J}\tau _{01j}F^{j})T$ and $%
(\sum_{j=0}^{J}\tau _{\pm 1j}F^{j})DT$, as Theorem 2 below shows in CRF2.
The interaction treatment takes the \textquotedblleft
natural\textquotedblright\ form $DT$ in CRF2, to which OLS\ can be easily
applied.\bigskip

\textbf{THEOREM 2.}\textit{\ Under Assumption 1, CRF1 simplifies to:}%
\begin{equation}
Y=\sum_{j=0}^{J}\beta _{0j}F^{j}\ +(\sum_{j=0}^{J}\delta _{0j}F^{j})D\
+(\sum_{j=0}^{J}\tau _{01j}F^{j})T\ +(\sum_{j=0}^{J}\tau _{\pm
1j}F^{j})\cdot DT+U.  \tag{CRF2}
\end{equation}

\qquad If order-2 approximation were used for $t$, then this would add $%
T^{2} $ and $DT^{2}$ terms in CRF2. One might argue that our CRF-based
estimation can also misspecify the functions of $F$ and $T$ as the
conventional T- and R-models can. This is true, but the functions of $F$ and 
$T$ in the CRF's are of RF varieties, not SF's as in the T- and R-models,
and the consequences of misspecifying RF's should be less serious than
misspecifying SF's. Our CRF-based estimation rests on this premise; i.e.,
specify RF's, not SF's, if one has to.

\section{Simulation Study}

\qquad Our simulation design uses the following as the basis: for an error
term $U_{3}$,%
\begin{eqnarray}
&&\ Y=\beta _{0}+\beta _{f}F\ +\ (\beta _{d}+\beta _{f2}\ln F)D\ +\ (\beta
_{\tau }+\beta _{r}F^{-1}+\beta _{f2}\ln F)T  \notag \\
&&\ \ \ \ \ \ \ \ \ \ \ \ \ \ \ \ \ \ \ \ \ \ +(\beta _{d\tau }+\beta
_{dr}F^{-1}+\beta _{f2}\ln F)DT\ +\ U_{3},  \notag \\
&&\ E(D)=0.5,\ \ \ U_{3}\sim N(0,1)\amalg (D,T,F),\ \ \ \beta _{0}=0,\ \beta
_{f}=-2,\ \beta _{d}=2,  \notag \\
&&\ \beta _{f2}=0,\ 0.4,\ \ \ \beta _{\tau }=0,\ 0.2,\ \ \ \beta _{r}=0,\
2,\ \ \ \beta _{d\tau }=0,\ 0.2,\ \ \ \beta _{dr}=0,\ 2,  \notag \\
&&\ N=1000,\ 2000\ \text{(and }5000\text{),}\ \ \ \ \ 1000\text{ repetitions.%
}  \TCItag{3.1}
\end{eqnarray}%
The parameter values for $\beta _{\tau }$ and $\beta _{d\tau }$ are set to
small numbers to make the network effect of all treated friends not too
different from the direct effect of $D$.

\qquad For each $i$, we generate its longitude and latitude from the
two-dimensional uniform distribution on $[0,1]^{2}$, and $F_{i}$ is the
number of units falling within the circle of radius $0.025$ from the unit $i$%
's location, and $T_{i}$ are the number of the treated among the $F_{i}$
friends. Then we select only those with $F>0$ to estimate the effects; about 
$63\%$ are retained when $N=1000$, and about $85\%$ when $N=2000$. Hence,
the selected sample size with $N=2000$ is about three times that with $%
N=1000 $.

\qquad In the selected sample with $N=1000$, we have $\bar{F}\simeq 2.5$
with $SD(F)\simeq 1.3$ where $SD$ stands for standard deviation, and $\bar{T}%
\simeq 1.6$ with $SD(T)\simeq 0.78$; the maximum value of $F$ is about $7$.
In the selected sample with $N=2000$, we have $\bar{F}\simeq 4.2$ with $%
SD(F)\simeq 1.9$, and $\bar{T}\simeq 2.3$ with $SD(T)\simeq 1.2$; the
maximum value of $F$ is about $12$. Whereas the effect of $D$ is always $%
\beta _{d}=2$, the network effect of $T$ and the interaction effect of $DT$
are%
\begin{equation*}
E(\beta _{\tau }+\beta _{r}F^{-1}+\beta _{f2}\ln F)\text{ \ \ \ \ and \ \ \
\ }E(\beta _{d\tau }+\beta _{dr}F^{-1}+\beta _{f2}\ln F).
\end{equation*}

\qquad Different parameter values result in different models: whereas the
CRF holds always,%
\begin{eqnarray}
&&\ (i)\text{\ }\beta _{\tau }=0.2\ \text{and }\beta _{f2}=\beta _{r}=\beta
_{d\tau }=\beta _{dr}=0:\text{T- and TR-models hold;}  \notag \\
&&\ (ii)\ \beta _{r}=2\ \text{and }\beta _{f2}=\beta _{\tau }=\beta _{d\tau
}=\beta _{dr}=0:\text{R- and TR-models hold;}  \notag \\
&&\ (iii)\ \beta _{f2}=0\text{ but }\beta _{\tau },\ \beta _{r},\ \beta
_{d\tau },\ \beta _{dr}\text{ are not zero }:\text{TR-model holds;} 
\TCItag{3.2} \\
&&\ (iv)\ \beta _{f2},\ \beta _{\tau },\ \beta _{r},\ \beta _{d\tau },\
\beta _{dr}\text{ are all not zero: no linear model holds.}  \notag
\end{eqnarray}

\qquad Table 1 presents the simulation results only for $N=2000$---the table
for $N=1000$ is omitted to save space: \textit{four OLS's to T-, R- and
TR-model, and CRF2 with quadratic approximations in }$F$. We consider four
data generating processes (DGP's) of (3.2) using the parameter values in
(3.1), as can be seen in the column headers of Table 1. The column `True
Effect' shows the true effects of $(D,T,DT)$ when each of the models
(i)-(iv) in (3.2) is correct; e.g., the first three rows in the True Effect
column are the true effects when (i) of (3.2) is correct, and the next three
rows are the true effects when (ii) is correct.

\qquad Since the true effects differ much across the DGP's, it might be
better to standardize the bias's and SD's by dividing them with the true
effects, which is not done, however, because some true effects are zero. For
each entry of the tables, the absolute bias and SD are presented. No SD
appears for the interaction effects in the T-OLS and R-OLS, because the
interaction effects are supposed (i.e., \textquotedblleft
estimated\textquotedblright ) to be zero.

\qquad In Table 1, first, when the DGP is the T model, the T-, TR- and
CRF-OLS's are unbiased, but R-OLS is biased by about 10\% (100$\times $%
0.02/0.20) in network effect as the network effect is wrong only in R-OLS.
Second, when the DGP is the R-model, the R- and TR-OLS's are mostly
unbiased, but T- and CRF-OLS's are biased in network effect. The bias of
30\% (=100$\times $0.18/0.61) is understandable for T-OLS as its network
effect is wrong, but the bias of about 15\% (100$\times $0.09/0.61) in
CRF-OLS is not small. Third, when the DGP is the TR-model, both T- and
R-OLS's are highly biased, whereas TR-OLS does best and CRF-OLS\ is somewhat
biased in three effects by about 6$\sim $13\%. Finally, when the DGP is the
CRF, the T-\ and R-OLS's are highly biased, and TR-OLS does still best
despite the misspecifications (the omitted $\ln F$), and CRF-OLS is biased
by 3$\sim $6\%.

\begin{center}
\begin{tabular}{lrccccc}
\hline\hline
\multicolumn{7}{c}{Table 1. $|$Bias$|$ \& SD of Four OLS's\ with $N=2000$}
\\ 
& \multicolumn{2}{l}{True Effect} & (i) T model & (ii) R model & (iii)\ TR & 
(iv)\ CRF \\ \hline
T-OLS & (i)$\ D$ & 2.00 & 0.00\ 0.05 & 0.00 0.05 & 1.60 0.06 & 2.95 0.09 \\ 
& $T$ & 0.20 & 0.00\ 0.03 & 0.18 0.03 & 0.15 0.04 & 0.57 0.05 \\ 
& $DT$ & 0.00 & 0.00\ ...... & 0.00 ...... & 0.80 ...... & 1.35\ ...... \\ 
R-OLS & (ii) $D$ & 2.00 & 0.00\ 0.05 & 0.00 0.05 & 1.60 0.06 & 2.95 0.10 \\ 
& $T$ & 0.61 & 0.02\ 0.03 & 0.00 0.04 & 0.45 0.04 & 0.91 0.08 \\ 
& $DT$ & 0.00 & 0.00\ ...... & 0.00 ...... & 0.80 ...... & 1.35\ ...... \\ 
TR-OLS & (iii) $D$ & 2.00 & 0.00\ 0.13 & 0.01 0.13 & 0.01 0.14 & 0.03 0.14
\\ 
& $T$ & 0.80 & 0.00\ 0.05 & 0.00 0.05 & 0.02 0.05 & 0.08 0.05 \\ 
& $DT$ & 0.80 & 0.00\ 0.06 & 0.00 0.07 & 0.01 0.07 & 0.02 0.07 \\ 
CRF-OLS & (iv) $D$ & 2.52 & 0.00\ 0.14 & 0.01 0.14 & 0.11 0.14 & 0.07 0.14
\\ 
& $T$ & 1.35 & 0.00\ 0.05 & 0.09 0.05 & 0.08 0.05 & 0.08 0.05 \\ 
& $DT$ & 1.35 & 0.00\ 0.07 & 0.00 0.07 & 0.09 0.07 & 0.07 0.07 \\ \hline
\multicolumn{7}{c}{`True Effect' is the true effects under the DGP's
(i)-(iv) in (3.2)} \\ \hline\hline
\end{tabular}
\end{center}

\qquad In Table 1, most results are what was expected for the four OLS's.
However, when the DGP is TR or CRF, TR-OLS performed better than expected
whereas CRF-OLS did worse. This might cast a doubt on CRF-OLS. To dissipate
the doubt, Table 2 with $N=5000$ presents part of the results using (iii)
and (iv) for TR-OLS and CRF-OLS. When the TR model is the DGP, TR-OLS does a
little better than CRF-OLS, with their SD's being almost the same. However,
when CRF is the DGP, CRF-OLS outperforms TR-OLS, with their SD's being the
same. These findings are remarkable, considering that CRF-OLS uses only
quadratic approximations---the most basic nonlinear extension from
linear---whereas the DGP has $F^{-1}$ or $(F^{-1},\ln F)$, not $F$ or $%
(F,F^{2})$.

\begin{center}
\begin{tabular}{lrccc}
\hline\hline
\multicolumn{5}{l}{Table 2. $|$Bias$|$ \& SD of Two OLS's with $N=5000$} \\ 
& \multicolumn{2}{l}{True Effect} & (iii) TR & (iv) CRF \\ \hline
TR-OLS & (iii) $D$ & 2.00 & 0.00 0.09 & 0.07 0.10 \\ 
& $T$ & 0.44 & 0.00 0.01 & 0.01 0.01 \\ 
& $DT$ & 0.44 & 0.00 0.02 & 0.00 0.02 \\ 
CRF-OLS & (iv) $D$ & 2.88 & 0.01 0.10 & 0.00 0.10 \\ 
& $T$ & 1.32 & 0.01 0.01 & 0.00 0.01 \\ 
& $DT$ & 1.32 & 0.01 0.02 & 0.00 0.02 \\ \hline\hline
\end{tabular}
\end{center}

\qquad In summary, despite good performances of the T- and R-model OLS in
some cases, overall, we find no good reason not to use the TR-model OLS,
which is almost as easy to implement as the T- and R-model OLS. The TR-model
can reveal which is correct between the T- and R-models, or neither. If the
sample size is large, CRF-OLS should be also applied, as it can reveal
whether the TR-model holds or not.

\qquad Why did TR-OLS work well in many cases? Recalling (2.2), the TR-model
specifies%
\begin{equation*}
E(Y^{00}|F)=\beta _{0}+\beta _{f}F,\text{ \ \ }\tau _{0t}(F)=(\beta _{\tau
}+\beta _{r}F^{-1})t,\ \ \ \tau _{\pm t}(F)=(\beta _{d\tau }+\beta
_{dr}F^{-1})t.
\end{equation*}%
These may be taken as linear specifications to the unknown functions of $F$,
as using $\beta _{r}F^{-1}$ is analogous to using $-\beta _{r}F$. Since
linear specifications often work well in practice, this seems to be why
TR-OLS performed well. CRF-OLS then used $F^{2}$ extra, whose benefits were
realized only in large samples, with multicollinearity problems appearing in
small samples. This explains the performances of CRF-OLS vis-\`{a}-vis
TR-OLS.

\section{Conclusions}

\qquad Network/spillover effect is an important issue, as human beings do
not live in isolation and are constantly affected through interactions with
others. Given a randomized binary treatment $D$, one popular model
(\textquotedblleft T model\textquotedblright ) used to find the network
effect takes the number of the treated friends $T$ as the network treatment,
and another popular model (\textquotedblleft R model\textquotedblright )
takes the ratio $T/F$ as the network treatment where $F$ is the number of
friends/neighbors. The two models are vastly different though: if a constant 
$\beta $ is the effect of $T$ in the T-model, then the effect of $T$ in the
R-model is $\beta /F$---heterogeneous in $F$. Hence, an important question
arises: which is the right model?

\qquad This paper provided an answer, considering a general model
(\textquotedblleft TR-model\textquotedblright ) that encompasses both
models. Estimating the TR-model with OLS, which is as easy as estimating the
T or R-model, we can see which is the right model, or neither is.

\qquad Since the effect heterogeneity in $F$ may not necessarily take the
form $\beta /F$, we further obtained a \textquotedblleft causal reduced form
(CRF)\textquotedblright\ that holds for any form of outcome $Y$ and for any
form of effect heterogeneity in $F$. The CRF contains causal parameters of
interest as slopes; e.g., denoting the potential outcomes as $Y^{dt}$ where $%
d$ and $t$ index the values of $(D,T)$, the effect $E(Y^{10}-Y^{00}|F)$ of $%
D $ with $T=0$ appears as the slope of $D$ in the CRF. The CRF can be
estimated by specifying the unknown functions such as $E(Y^{10}-Y^{00}|F)$.
Of course, misspecifications can occur, but the functions such as $%
E(Y^{10}-Y^{00}|F)$ are \textquotedblleft reduced forms
(RF's)\textquotedblright , not \textquotedblleft structural forms
(SF's)\textquotedblright\ as the T- and R-models are, and specifying RF's
should be less riskier than specifying SF's.

\qquad In our simulation study, the TR-model worked well, and only in large
samples, OLS for CRF outperformed OLS\ for the TR-model. This was attributed
to that the TR-model essentially uses a linear approximation to unknown
functions such as $E(Y^{10}-Y^{00}|F)$, and OLS\ for CRF uses power
functions of $F$ extra. Given that the TR-model was motivated by the simple
desire to encompass both T- and R-models, finding its good performance is
something of a \textquotedblleft serendipity\textquotedblright . Our hope is
that researchers see the advantages of the TR-model and CRF-based
estimation, and apply them fruitfully in their research.\bigskip 

\begin{center}
{\LARGE Appendix:\ Proof of Theorem 1\medskip }
\end{center}

\qquad Using $Y^{dT}=\sum_{t=0}^{F}1[T=t]\cdot Y^{dt}$, $%
Y=(1-D)Y^{0T}+DY^{1T}$ becomes%
\begin{eqnarray*}
Y &=&(1-D)\sum_{t=0}^{F}1[T=t]\cdot Y^{0t}\ +\ D\sum_{t=0}^{F}1[T=t]\cdot
Y^{1t} \\
&=&(1-D)\{Y^{00}+\sum_{t=1}^{F}1[T=t](Y^{0t}-Y^{00})\}+D\{Y^{10}+%
\sum_{t=1}^{F}1[T=t](Y^{1t}-Y^{10})\} \\
&=&Y^{00}+(Y^{10}-Y^{00})D\ +\
\sum_{t=1}^{F}\{(1-D)(Y^{0t}-Y^{00})+D(Y^{1t}-Y^{10})\}\cdot 1[T=t] \\
&=&Y^{00}+(Y^{10}-Y^{00})D\ +\ \sum_{t=1}^{F}(Y^{0t}-Y^{00}+\Delta
Y_{t}^{\pm }D)\cdot 1[T=t].
\end{eqnarray*}

\qquad Since $T$ is as good as randomized given $F>0$, $%
E(Y^{dt}|D,T,F)=E(Y^{dt}|F)$ holds when $F>0$. Also, $F=0$ implies $T=0$ to
drop $T$ from $E(Y^{dt}|D,T,F=0)$. Hence, $E(Y^{dt}|D,T,F)=E(Y^{dt}|F)$
holds always. Take $E(\cdot |D,T,F)$ on the above $Y$ equation:%
\begin{equation*}
E(Y|D,T,F)=E(Y^{00}|F)+E(Y^{10}-Y^{00}|F)D+\sum_{t=1}^{F}\{\tau
_{0t}(F)+E(\Delta Y_{t}^{\pm }|F)D\}\cdot 1[T=t].
\end{equation*}%
Then, $U\equiv Y-E(Y|D,T,F)$ renders CRF1 with $E(U|D,T,F)=0$. Also,%
\begin{eqnarray*}
\tau _{1t}(F)-\tau _{0t}(F)
&=&E(Y^{1t}-Y^{10}|F)-E(Y^{0t}-Y^{00}|F)=E(\Delta Y_{t}^{\pm }|F); \\
\delta _{t}(F)-\delta _{0}(F)
&=&E(Y^{1t}-Y^{0t}|F)-E(Y^{10}-Y^{00}|F)=E(\Delta Y_{t}^{\pm }|F).
\end{eqnarray*}

\begin{center}
{\LARGE REFERENCES}
\end{center}

\qquad Aronow, P.M. and C. Samii, 2017, Estimating average causal effects
under general interference, Annals of Applied Statistics 11, 1912-1947.

\qquad Bryan, G., S. Chowdhury and A.M. Mobarak, 2014, Underinvestment in a
profitable technology: the case of seasonal migration in Bangladesh,
Econometrica 82, 1671-1748.

\qquad Cai, J., A. De Janvry and E. Sadoulet, 2015, Social networks and the
decision to insure, American Economic Journal: Applied Economics 7, 81-108.

\qquad Choi, J.Y., G. Lee and M.J. Lee, 2023, Endogenous treatment effect
for any response conditional on control propensity score, Statistics and
Probability Letters 196, 109747.

\qquad Choi, J.Y. and M.J. Lee, 2018, Regression discontinuity with multiple
running variables allowing partial effects, Political Analysis 26, 258-274.

\qquad Forastiere, L., E.M. Airoldi and F. Mealli, 2021, Identification and
estimation of treatment and interference effects in observational studies on
networks, Journal of the American Statistical Association 116, 901-918.

\qquad Hu, Y., S. Li and S. Wager, 2022, Average direct and indirect causal
effects under interference, Biometrika 109, 1165-1172.

\qquad Kim, B.R., 2025, Estimating spillover effects in the presence of
isolated nodes, Spatial Economic Analysis, forthcoming.

\qquad Kim, B.R. and M.J. Lee, 2024, Instrument-residual estimator for
multi-valued instruments under full monotonicity, Statistics and Probability
Letters 213, 110187.

\qquad Lee, G., J.Y. Choi and M.J. Lee, 2023, Minimally capturing
heterogeneous complier effect of endogenous treatment for any outcome
variable, Journal of Causal Inference 11, 20220036.

\qquad Lee, M.J., 2018, Simple least squares estimator for treatment effects
using propensity score residuals, Biometrika 105, 149-164.

\qquad Lee, M.J., 2021, Instrument residual estimator for any response
variable with endogenous binary treatment, Journal of the Royal Statistical
Society (Series B) 83,\ 612-635.

\qquad Lee, M.J., 2024, Direct, indirect and interaction effects based on
principal stratification with a binary mediator, Journal of Causal Inference
12, 20230025\textit{.}

\qquad Lee, M.J. and C. Han, 2024, Ordinary least squares and
instrumental-variables estimators for any outcome and heterogeneity, Stata
Journal 24, 72-92.

\qquad Lee, M.J., G. Lee and J.Y. Choi, 2025,\ Linear probability model
revisited: why it works and how it should be specified, Sociological Methods
\& Research 54, 173-186.

\qquad Leung, M.P., 2020, Treatment and spillover effects under network
interference, Review of Economics and Statistics 102, 368-380.

\qquad Manski, C.F., 2013, Identification of treatment response with social
interactions, Econometrics Journal 16, S1-S23.

\qquad Mao, H. and L. Li, 2020, Flexible regression approach to propensity
score analysis and its relationship with watching and weighting, Statistics
in Medicine 39,\ 2017-2034.

\qquad Miguel, E. and M. Kremer, 2004, Worms: identifying impacts on
education and health in the presence of treatment externalities,
Econometrica 72, 159-217.

\qquad Oster, E. and R. Thornton, 2012, Determinants of technology adoption:
peer effects in menstrual cup take-up, Journal of the European Economic
Association 10, 1263-1293.

\qquad Van der Laan, M.J. and S. Rose, 2011, Targeted learning: causal
inference for observational and experimental data, Springer.

\end{document}